\documentclass[twocolumn]{aastex63}
\usepackage{lineno}
\usepackage{xcolor}
\usepackage{graphicx, natbib, color, bm, amsmath, epsfig, mathrsfs, xfrac, xspace}
\usepackage[inline]{enumitem}
\usepackage{booktabs}
\usepackage{ifthen}
\usepackage{ulem}
\usepackage{acronym}

\usepackage{hyperref}
\usepackage{float}

\usepackage{rotating}
\usepackage{ulem}
\definecolor{orange}{rgb}{1.        ,  0.54,  0}
\definecolor{meta}{rgb}{0.371,0.617,0.625} 
\definecolor{redo}{rgb}{1.,0.,0.}  
\newcommand{\newtextc}[1]{{ #1}}

\def\nar{New Astronomy Reviews}

\def\hydra{\textit{HYDRA}}
\begin{document}
\title{Galactic Positrons from Thermonuclear Supernovae}
\author[0000-0001-5888-2542]{T. B. Mera Evans}
\affil{Department of Physics, Florida State University, Tallahassee, Fl 32306, USA}
\author[0000-0002-4338-6586]{P. Hoeflich}
\affiliation{Department of Physics, Florida State University, Tallahassee, Fl 32306, USA}
\author[0000-0003-2682-3067]{R. Diehl}
\affil{Max Planck Institut f\"ur extraterrestische Physik, D-85748 Garching, Germany}

\begin{abstract}
Type Ia Supernovae (SNe~Ia) may originate from a wide variety of explosion scenarios and progenitor channels. They exhibit a factor of $\approx 10$ difference in brightness and, thus, a differentiation in the mass of $^{56}$Ni$ \rightarrow ^{56}$Co $\rightarrow ^{56}$Fe. We present a study on the fate of positrons within SNe~Ia in order to evaluate their escape fractions and energy spectra. 
Our detailed Monte Carlo transport simulations for positrons and $\gamma $-rays include both $\beta ^+$ decay of $^{56}$Co and pair production. 
We simulate a wide variety of explosion scenarios, including the explosion of white dwarfs (WD) close to the Chandrasekhar mass ($M_{\mathrm{Ch}}$), He-triggered explosions of sub-$M_{\mathrm{Ch}}$ WDs, and dynamical mergers of two WDs. 
For each model, we study the influence of the size and morphology of the progenitor magnetic field between 1 and $10^{13}$ G. 
\newtextc{
Population synthesis based on the observed brightness distribution of SNe~Ia was used to estimate the overall contributions to Galactic positrons due to escape from SN Ia. We find that this is dominated by normal-bright SNe~Ia, where variations in the distribution of emitted positrons are small.
We estimate a total SNe~Ia contribution to the Galactic positrons of $< 2\% $ and, depending on the magnetic field morphology, $<6...20\%$ for $M_{\mathrm{Ch}}$ and sub-$M_{\mathrm{Ch}}$,
respectively. }


\end{abstract}

\section{\label{sec:Introduction}INTRODUCTION}


The annihilation of positrons produces a characteristic gamma-ray spectrum with a prominent line at 511 keV.
This characteristic emission has first been discovered in solar flare measurements, and has been observed from our Galaxy since the 1970s \citep{haymes_1974}\citep[see][for a review]{prantzos_2011}. After an early period of reports of variability and the possible roles of high-energy transient sources, the diffuse and steady nature of this gamma-ray emission was established by long-term monitoring with the Ramaty Solar Maximum Mission (SMM) \citep{share_1988}. Observations by the Oriented Scintillation Spectrometer Experiment (OSSE) were able to obtain coarse imaging, and showed the emission to be extended along the inner Galactic plane, with prominence of emission regions in the central few kpc of the Galaxy \citep{purcell_1997}. More recently, the gamma-ray spectrometer SPI aboard the International Gamma-Ray Astrophysics Laboratory (INTEGRAL) obtained improved imaging \citep{knoedlseder_2005}, detailed spectral information \citep{jean_2006}, and high-quality spectra that were even resolved for different emission regions along the Galaxy, i.e. the outer disk, the inner disk and bulge, and the central region with the supermassive black hole SgrA \citep{siegert_2016} \citep[see also][for a recent observational review]{churazov_2020}. 
The astrophysical puzzle from all these observations and theoretical studies on Galactic positron origins is that among the many possible sources that could produce a large supply of positrons, either the amount of positrons or their spatial distribution is at odds with observational constraints \citep[see][for a review]{prantzos_2011}.

Cosmic rays, SNe~Ia, core collapse supernovae, pulsars, dark matter annihilation, and X-ray binaries are among some of the possible sources, each of which may produce positrons with energies of $\approx 1$ MeV, \newtextc{an energy low enough for positrons to be confined in the Galactic disk} 
\citep{prantzos_2011, Siegert2017PhD}. 
To discriminate between possible sources, it was often assumed that positrons would annihilate locally, so that the observed emission could be attributed directly to their spatial distribution.
Studies have consistently shown that the emission morphology differs significantly from the spatial morphology of each of the candidate sources, which has lead to increased studies on interstellar positron propagation  \citep{jean_2009, alexis_2014}. From these, positrons may propagate through interstellar space over kpc regions, where they have lost energies through scatterings and inelastic collisions to energies less than $\approx 1$ keV. At this point, they can then form positronium, annihilate, and shape the observed characteristic gamma-ray spectrum.
Galactic positron propagation simulations from nucleosynthesis origins have been made, in particular modelling the spatial morphologies expected from supernovae of all scenarios \citep{martin_2012, alexis_2014, higdon_2009}. Evidently, the observed emission \citep{siegert_2016} does not agree with such simulations.
This issue has collectively become known as the positron puzzle.

The role of SNe~Ia had been emphasized early on \newtextc{(\cite{skibo_1992}, see \cite{prantzos_2011} for a review)}: the abundantly-produced $^{56}$Ni from these sources and its decay could explain the total Galactic flux of annihilation emission, with plausible assumptions on positron leakage, given that each SN~Ia would produce about $0.6 M_{\odot}$ of this isotope.

\newtextc{Considering the uncertainties of SNe~Ia locations and of positron propagation in interstellar medium around the sources, constraints from the spatial distribution of observed annihilation emission are difficult.}
\newtextc{Better} constraints may be feasible, but only once the absolute amount of positron injections from SNe~Ia is specified more precisely. 
This is the objective of our study. We want to make use of the latest tools of positron propagation within SNe~Ia and explore the variety of explosion scenarios, in the hope that we can constrain the contribution from these types of sources.

 The consensus is that SNe~Ia result from the thermonuclear runaway of a degenerate C/O white dwarf (WD) from a close binary stellar system \citep{Hoyle1960}. From the \emph{quasi-standard candle} nature of SNe~Ia light curves
 \citep{Phillips1993} to the growing observational spectral differences between SNe~Ia \citep{quimby06, Polin19}, it has increasingly become clear that there is a diverse pool of SNe~Ia, which may indicate variations in progenitor channels and explosion scenarios \citep{branch93, pakamor11, hoeflich2013, ckjt13, gm13}. 
 
 There are two main potential progenitor systems that either consists of two WDs, a double-degenerate system (DD), or a single WD and a main-sequence, helium, or red giant star, a so-called single-degenerate system (SD) \citep{ibentut84, webbink84, han06a, Di_Stefano2011}. 
 
 The explosion scenario is a highly debated topic, but there are three leading mechanisms. In the first scenario, the explosion is triggered by compressional heating near the center of an accreting  WD as it approaches the Chandrasekhar mass limit ($M_{\mathrm{Ch}}\approx1.4M_{\odot}$), in either a DD or SD system. The flame starts as a deflagration \citep{nomoto84}, and most likely transitions to a detonation (deflagration to detonation transition (DDT)) \citep{khok89,hk96,ww91,yamaoka92}. In the second scenario, a surface layer of He becomes critical and detonates, and thus triggers a secondary central detonation in the C/O core  of a sub-$M_{\mathrm{Ch}}$ WD \citep{wwt80, nomoto82, livne1990, hk96, kromer10, sim10,WK2011, Shen2015, Tanikawa2018, Glasner2018}. For the third scenario, merging or colliding WDs can trigger an explosion from the heat release caused by their dynamical time scales \citep{benz90,rasio94,hk96,segretain97,yoon2007,WMC09,WCMH09,loren09,Pakmor10,isern11,pakmor12,Rosswog2009,Thompson2011,Pejcha2013,Kushnir2013,Dong2015}.
 
  In this paper, our first goal is to present \emph{positron escape fractions} and \emph{positron energy distributions} for detailed positron and $\gamma$-ray transport solutions for the wide range of SN~Ia explosion scenarios, encompassing subluminous, transitional, and normal bright SNe~Ia that test dipole and small scale turbulent magnetic field configurations that range in strength from 1 to $10^{13}$ G. The second goal is to then analyze the possible \emph{impact SNe~Ia will have on the total Galactic positron budget and annihilation rate}, by examining the observed population synthesis of SNe~Ia as a function of brightness where we will then evaluate this result with the SN~Ia Galactic rate. 
 
Thus, we organize our paper as follows: $\S$ \ref{sect:moti} provides context behind our motivation, $\S$ \ref{sect:esc_frac} discusses the escape fractions and energy spectrums from our detailed simulations for a wide variety of SN~Ia explosion models and explores case studies of various explosion models weighted by the observed ratio of SN~Ia classes, $\S$ \ref{sect:con} discusses our results in the context of previous studies and explores the possible impact SNe~Ia may have on the Galactic 511 keV emission feature, and finally, we conclude on our study.

\section{Motivation}
\label{sect:moti}

 The bolometric and monochromatic light curves and spectra are powered by the radioactive decays of $^{56}$Ni $\rightarrow ^{56}$Co $\rightarrow ^{56}$Fe on time scales of 6.1 and 111.4 days, respectively, which heats the envelope and produces the photons we observed as a supernova over the course of months to years. 
 A most-direct observational proof of this was obtained when SN~Ia SN2014J occurred at a relatively nearby distance, so that characteristic gamma rays from the decays of $^{56}$Ni and $^{56}$Co could be observed \citep{Churazov:2014,Diehl:2014}, together with its optical light curve and spectra that identified it as a SN~Ia, and the nebular spectral evolution from a Co to an Fe dominated phase.
 The production and distribution of $^{56}$Ni depends on the explosion scenario and the flame physics \citep{Hoeflich2017book}. $^{56}$Co decays via electron capture and positron emission. Depending on the properties of the envelope and magnetic field in the SN~Ia ejecta, the positrons may deposit their energy in the envelope and/or may escape and contribute to the total amount of positrons in our Galaxy.
 Gamma-ray spectra from SN2014J as measured with INTEGRAL clearly showed the characteristic line at 511~keV, that signified that positron annihilation within the supernova indeed occurred \citep{Churazov:2014}.
 The intensity of the observed line is consistent with annihilation of most of the positrons from $^{56}$Ni decay; within uncertainties, only a minority of positrons (at the percent level) would escape into the surrounding medium \citep{Sonnberger:2019}. 

This leads to a basic question: What is the fraction of escaping positrons from SNe~Ia, and what is their resulting energy distribution?

\citet{chan_ligenfelter_1993} were pioneers in studying the effects of magnetic fields on positron escape fractions and energy distributions from four SN~Ia models, where they assumed either a radial magnetic field or a locally trapping turbulent structure.  
\citet{milne_1999_true, milne_2001} extended this approach by implementing a positron transport Monte Carlo scheme for a larger range of SN~Ia models with similar magnetic field structures. 

\newtextc{\citet{cappellaro_1997} suggested that optical light curves can be used as a proxy for the bolometric light curve, and that the positron escape can be measured by the deviation of observed brightness from the total energy available from $^{56}$Co decay.
Following this approach and because the light curves declined faster than the radioactive decay already some 80 to 150 days past the explosion, \citet{milne_1999_true, milne_2001} concluded that the early positron escape strongly favored radial B-fields. Because of the expected redistribution from the optical to the IR \citep{fransson94}, their approach that optical light curves trace bolometeric energy deposition was drawn into question by \citet{stritzinger02} and \citet{Sollerman2004}, who included NIR observations and found consistency with complete positron trapping. The importance of photon redistribution was confirmed in many subsequent studies \citep{stritzinger_2007,leloudas_2009,kerzendorf_2014,Gall2018}.}


Recent simulations of positron transport within SNe~Ia \citep{penney14} and additional analysis of light curves and spectral profiles \citep{hoeflich2013, tiara15, Diamond2018, hristov_2021} have shown the need for large initial fields and either small scale turbulent or dipole magnetic field morphologies.

\newtextc{\citet{hristov_2021} reconstructed bolometric light curves based on data in literature (\citet{cappellaro_1997, leloudas_2009, kerzendorf_2014, graur_2016, yang2018}, and references therein). 
To determine the deviation from the total radioactive energy, they 
included contributions in the NIR and MIR, and provided upper limits for the positron dominated tail
(see their Fig. 11). 
Where no NIR or MIR data were available, the redistribution was taken from normal bright and transitional SNe models \citep{hoeflich2017, Gall2018} and from the redistribution functions of SN2014J \citep{Telesco2015}.
For the estimates, \citet{hristov_2021} assumed that positron escape would
cause the bolometric light curve to fall below the $^{56}$Co-decay line, whereas additional energy sources, such as
 $^{57}$Co seen at later times \citep{graur_2016, yang2018}, would increase the brightness, thus limiting the useful range of constraining the magnetic field in the epoch between 1 and 3 years after maximum light.
All light curves were consistent with no positron escape, but the uncertainties in their reconstruction are responsible for lower magnetic field limits from both the light curves and line profiles, which are $\approx 10^{6}$ and $10^{9}$ G for turbulent and dipole fields, respectively \citep{penney14,Diamond2015,Diamond2018,hristov_2021}.
In spite of these interpretational discussions, these works did not address positron escape fractions and energy distributions,} and were limited to specific explosion models.

\section{Positrons from Our SN~Ia models}
\label{sect:esc_frac}

The simulations and the explosion models used in this work utilize our HYDrodynamical RAdiation code (\hydra ) \citep{h90,h95,hoeflich2003hydra,hoeflich_2009_hyro,penney14, hristov_2021}.
Our simulations include both positron and $\gamma$-ray transport. More specifically, we include the positrons from the $\beta ^{+}$ decay channel of $^{56}$Co, but we also take into account the positron generation due to pair production of the $\gamma $-rays produced.

\begin{figure}[htb]
    \centering
    \includegraphics[width=\linewidth]{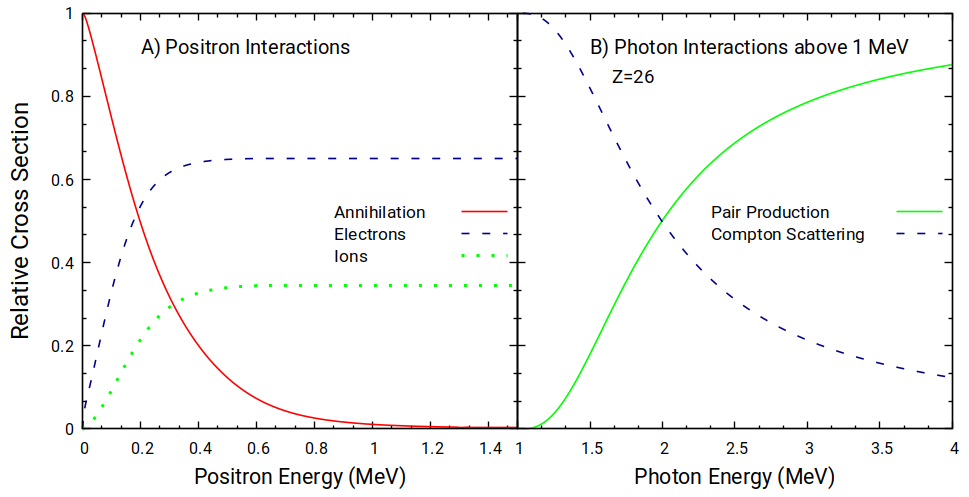}
    \caption{\newtextc{Left: Relative strength of cross sections of positron interactions for free electron in a plasma, a single ionization, and annihilation as a function of energy. Right: Relative strength of cross sections of photon interactions for pair production and Compton scattering as a function of energy for energies greater than 1 MeV and Z=26.}}
    \label{cross_sec_strength}
\end{figure}

The $\gamma$-ray and positron transport uses a module developed by \citet{penney14} that is based on the photon transport Monte-Carlo scheme from \citet{h02gam}. This approach follows \citet{amba88}, for which Compton scattering, ionization, bremsstrahlung, synchrotron radiation and pair production effects are included. We also take into account three dominant processes of interactions: atomic electron scatterings as positrons traverse through the media, excitations of free electrons in a plasma, and direct annihilations with electrons (where the cross sections and energy loss rates can be found in \citet{chan_ligenfelter_1993, lang_1999}; and \cite{gould_1972}). The relative strength of the cross sections can be found in Figure \ref{cross_sec_strength}.

The positron energy spectrum used in the initial injection is given by \citep{nadyozhin_1994}:
\begin{equation}
    N(E) = C p^2 (E_{0}-E)^2 (2 \pi \eta (1-exp(-2\pi \eta))^{-1})
\end{equation}
where $C$ is a scale factor and $\eta$ is the charge of the nucleus times $\hbar$ over the velocity of the electron. We also take into account energy from the pair production by high energy $\gamma$-rays from $^{56}$Ni decay. \newtextc{ The cross-section for this channel of positron production has been discussed by \citet{amba88}, where the relative strength of the cross sections for energies above 1 MeV can be found in Figure \ref{cross_sec_strength}.
\newtextc {As most $\gamma$-rays escape without interaction at late times, pair production contributes only a few percent
to the total positron production after $\approx 100$ days \citep{penney14}, and goes down at 200d and 500d to about 0.4 \% and 0.2\%, respectively. Note that those positrons may be deposited at the outer layers above the $^{56}$Ni layers with a high positron escape probability.}}\footnote{\newtextc{For computational efficiency,
we did not collect the branching ratio between pair production and total cross-section but estimated it by convolving the high-energy gamma spectrum with the total energy deposition, and found pair-production between $\approx$ 0.1 to 1 \%.}}

 \begin{table*}[hbt!]
\begin{center}
    
\caption{\newtextc{Time-integrated} positron escape fraction for each model in \%. Column one indicates the model and column two gives the respective total amount of positrons produced . The preceding columns then give the percentage of escaped positrons for various dipole or turbulent magnetic field strengths. The nomenclature used can be found in $\S$ \ref{sect:esc_frac}.}
\newtextc{\begin{tabular}{c|c|cccccccc}
\hline
\hline
 & Total & \multicolumn{2}{c}{B0} & \multicolumn{2}{c}{B3} & \multicolumn{2}{c}{B6} & \multicolumn{2}{c}{B9} \\ 
 Model& $e^{+}s$ & dipole & turb & dipole & turb & dipole & turb & dipole & turb \\ 
 &($10^{54}$)&(\%)&(\%)&(\%)&(\%)&(\%)&(\%)&(\%)&(\%)\\\hline
DDs05 & 0.94 & 0.23 & 0.23 & 0.23 & 0.23 & 0.05 & 0.04 & 0.04 &   $<$ 0.01 \\
DDs20 & 0.48 & 0.36 & 0.36 & 0.36 & 0.36 & 0.08 &  $<$ 0.01 & 0.07 &  $<$ 0.01\\
DDt20 & 1.19 & 1.61 & 1.61 & 1.61 & 1.61 & 0.52 & 0.06 & 0.50 & 0.03 \\
DDn11 & 2.57 & 2.88 & 2.80 & 2.88 & 2.80 & 1.09 & 0.17 & 1.08 & 0.09 \\
DDn20 & 2.49 & 3.05 & 2.97 & 3.05 & 2.97 & 1.14 & 0.18 & 1.12 & 0.09  \\
DDn40 & 2.39 & 3.09 & 3.02 & 3.09 & 3.02 & 1.19 & 0.19 & 1.17 & 0.10 \\
DET2 & 2.49 & 5.99 & 5.97 & 5.99 & 5.97 & 4.91 & 0.81 & 3.36 & 0.39  \\
HED8 & 2.01 & 2.60 & 2.60 & 2.60 & 2.60 & 0.91 & 0.05 & 0.87 & 0.02  \\
MWD & 2.37 & 4.59 & 5.33 & 4.59 & 5.33 & 3.24 & 1.43 & 2.57 & 0.86  \\
CO09 & 0.70 & 1.92 & 1.92 & 1.92 & 1.92 & 0.66 & 0.11 & 0.63 & 0.06  \\ \hline
\end{tabular}}
\label{tab:tab2}
\end{center}
\end{table*}
 
\subsection{Magnetic Field Limits}
The gyro motion of the positrons is significantly influenced by the size of the Larmor radius relative to the morphology of the B-field and scale variations of the explosion model, such as the density scale height. 

In the small B-field limit, where the Larmor radius is much larger than the structures of the explosion model, the path of the positron is nearly straight and can be integrated along linear rays. This was the approach taken by \citet{chan_ligenfelter_1993} and \citet{milne_1999_true, milne_2001} for their radial field assumptions. However, there also exists large and intermediate magnetic fields, where the Larmor radius is either much smaller or approximate to the structures in the model. In these regimes, the path of the positron needs to be explicitly integrated by dissecting the path length into sufficiently small segments so that the B-field changes along each segment are tracked accurately. 

A more thorough discussion of the three cases is found in \citet{penney14} \newtextc{(and their Table 1). They showed that, a few hundred days after the explosion, the positrons travel along linear rays for B-fields less than $10^3$ G, whereas positrons travel along the magnetic field lines for B-fields well in excess of $10^6$ G.}
 They also demonstrate that turbulent B-fields require full transport effects for any B-field strength; this differs from the assumption by \citet{chan_ligenfelter_1993} and \citet{milne_1999_true} that the positrons would be trapped and carried by the local material.   
 
\begin{figure}[htb]
    \centering
    \includegraphics[width=0.7\linewidth]{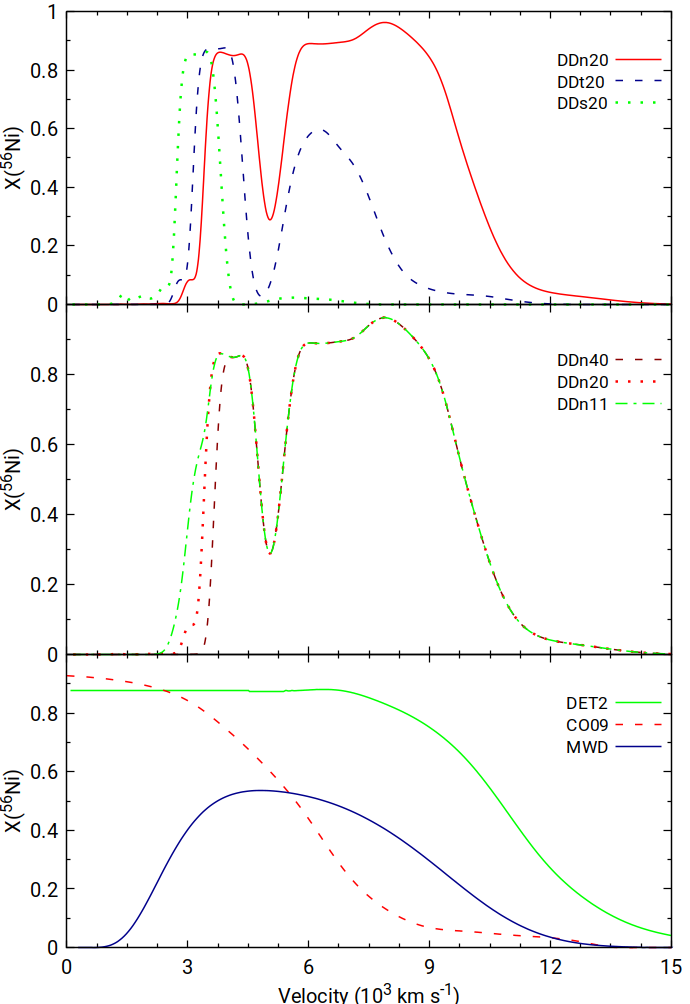}
    \caption{\newtextc{Abundance of $^{56}$Ni at $t \approx 0$ days after the explosion as a function of expansion velocity. Top: difference in total $^{56}$Ni mass for DDT models. Middle: difference in $^{56}$Ni mass for normal bright DDT models with different central densities. Bottom: Difference in $^{56}$Ni mass for various He triggered models and the merging scenario. The nomenclature used is described in $\S$ \ref{sect:esc_frac}.} }
    \label{ni_dist}
\end{figure}

\subsection{Model Nomenclature}
A more thorough summary of our SN~Ia explosion models and detailed structures can be found in \citet{HGFS99by02,hk96,Diamond2015,hoeflich2017}, and in Table \ref{table_mod}. \newtextc{For illustration, the $^{56}$Ni distributions of a subset are shown in Figure \ref{ni_dist}.\footnote{\newtextc{Detailed model structures are available on request.}}} The nomenclature for each model in this paper may differ from
literature. We present three different explosion scenarios for different magnetic field strengths and configurations, as described above. Each figure will be labeled to differentiate between magnetic field configurations, but to determine the magnetic field strength, the number after B"X" is the logarithmic strength in Gauss. MWD stands for merging white dwarfs. For DDT models we label them with a 'DD', the ensuing
letter, 'n', 't', and 's', stand for normal bright, transitional and subluminous respectively, and the final number stands for the central density in $10^{8}$ g~cm$^{-3}$. \newtextc{For helium detonations in sub-$M_{\mathrm{Ch}}$ models, 
HeD's explosion is triggered by a massive He-shell, whereas CO09 and DET2
are centrally triggered C/O cores. These are often used as proxy for surface He-triggered detonations (e.g. \citet{Blondin2017}), where the low mass He-shells avoid the production of surface $^{56}$Ni not seen in early time spectra.}

\begin{figure}[htb]
    \centering
    \includegraphics[width=\linewidth]{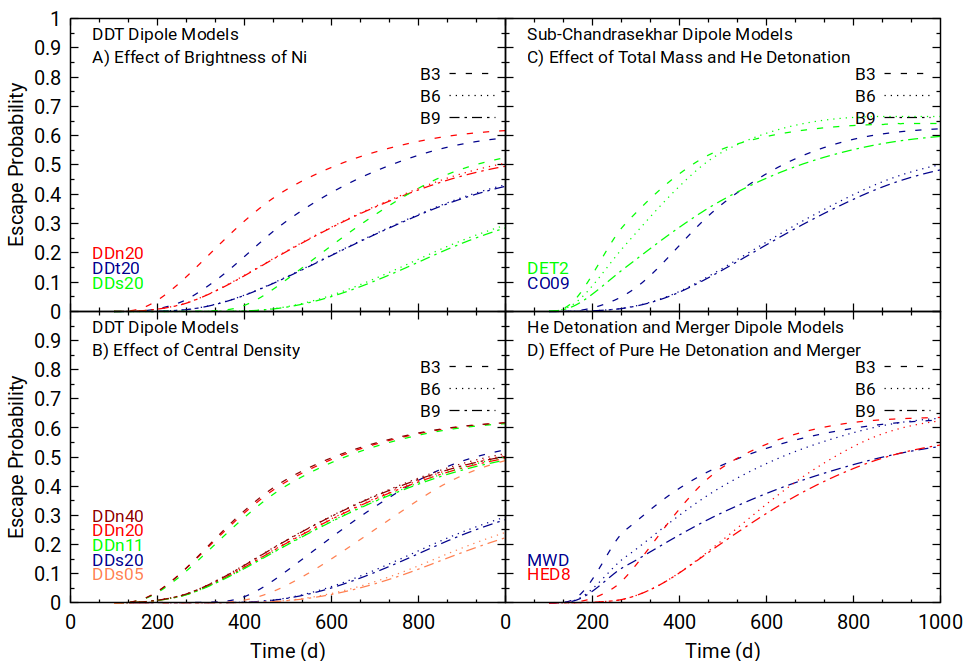}
    \caption{\newtextc{Escape probability of positron as a function of time} for all dipole models.  Panel A shows difference between total $^{56}$Ni production (upper left), panel B shows difference between central density (lower left), panel C shows difference between helium detonation masses (upper right), and panel D shows the difference between a merger and a helium detonation (lower right). The nomenclature used is described in $\S$ \ref{sect:esc_frac}. }
    \label{esc_frac}
    \centering
    \includegraphics[width=\linewidth]{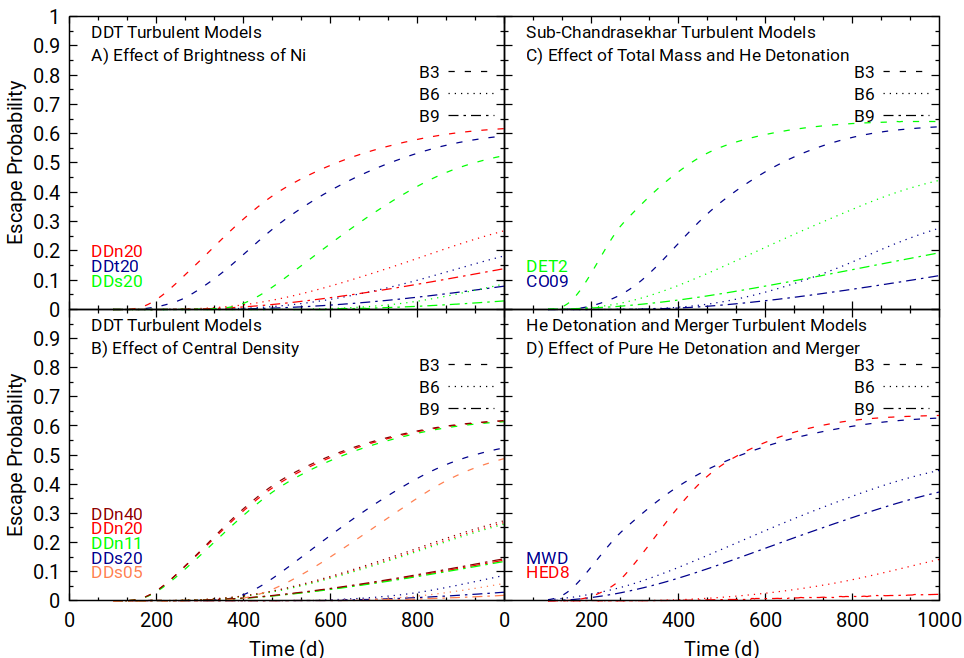}
    \caption{Same as Figure \ref{esc_frac}, but with turbulent models. \newtextc{Note that the highest B-fields are close to the time-axis.}}
    \label{esc_frac_turb}
\end{figure}

\subsection{Escape Fractions}
\label{test}

We show our escape study results in \newtextc{Figures 3-8}.
For these  figures, we group the models to emphasize differences in: the total amount of $^{56}$Ni production for DDT models with similar central densities (panel A in the upper left corner of Figures \newtextc{3-8}), the effect of different central densities for subluminous and normal bright DDT models (panel B in the lower left corner of Figures \newtextc{3-8}), the effect of total mass for helium detonations (panel C in the upper right of Figures 1-6), and the difference between a pure helium detonation and DD scenario (panel D in the lower right of Figures \newtextc{3-8}). \newtextc{For each model, the results for B0 are similar to those of B3 because such low B-fields hardly affect the path of a positron. B6 is close to the transition from low to high magnetic fields. For high magnetic fields ($\geq$ B9),
only those positron escape after being aligned with the magnetic field by multiple scattering, which result in similar escape probabilities, but incur a marginal shift in the 
energy distribution.}

Figures \ref{esc_frac} and \ref{esc_frac_turb} show the \newtextc{escape probability of} positrons from day 0 to day 1000, for dipole and turbulent models, respectively. Initial densities in the ejecta are high, and fall off exponentially as the edge of the envelope is approached. Homologous expansion reduces the density with time $t$ as $1/t^{3}$. However, the densities are such that escape does not occur until about day 200 for normal bright, helium, and merging explosion scenarios, with dipole magnetic fields $\leq 10^{3}$ G. The late initial escape for the dipole models may also be attributed to the positrons' interactions until directional alignment is found amongst the dipole magnetic field lines, so that they eventually escape at the poles. 

\begin{figure}[t]
    \centering
    \includegraphics[width=\linewidth]{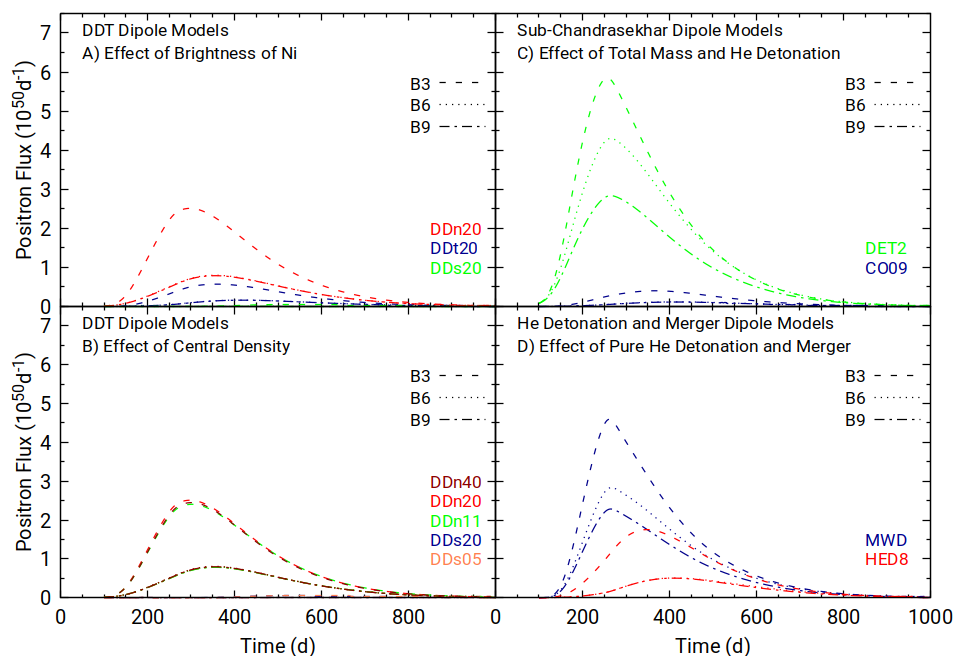}
    \caption{\newtextc{Positron flux} per day from day 0 to day 1000 for all dipole models. Panel A shows difference between total $^{56}$Ni production (upper left), panel B shows difference between central density (lower left), panel C shows difference between helium detonation masses (upper right), and panel D shows the difference between a merger and a helium detonation (lower right). The nomenclature used is described in $\S$ \ref{sect:esc_frac}.}
    \label{flux_day_dipol}
    \centering
    \includegraphics[width=\linewidth]{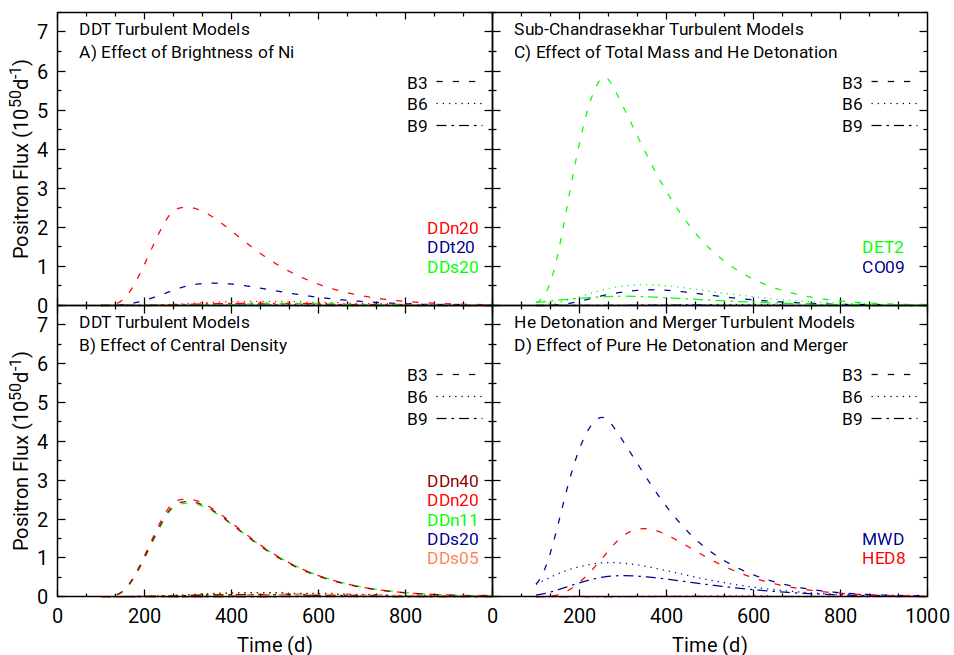}
    \caption{Same as Figure \ref{flux_day_dipol}, but with turbulent models. \newtextc{Note that the highest B-fields are close to the time-axis.}}
    \label{flux_day_turb}
\end{figure}

This delayed initial escape is further shifted to about day 300 for dipole magnetic fields $\geq 10^{6}$ G, which is due to the increased path length and longer diffusion time for positrons, as they gyrate around the magnetic field lines more tightly. This increased path length also contributes (with  a consistent factor of approximately two) to the difference between escape probabilities for magnetic fields $\leq 10^{3}$ G and $\geq 10^{6}$ G, as the positrons have a higher chance to annihilate in the envelope. The shift to day 400 and 500 for subluminous models is due to $^{56}$Ni production being produced at lower velocities \newtextc{(see Figure \ref{ni_dist}).}

These trends are also found in turbulent models, but with a continuous drop in escape probability for B-fields $\geq 10^{6}$ G, where escape begins to occur at around day 400. The additional path length along the turbulent magnetic field reduces escape fractions by an additional factor of two for these higher B-fields. The marginal difference between escape fractions for dipole and turbulent fields $\leq 10^{3}$ G is due to the size of the Larmor radius. By day 100, the Larmor radius is already 1.9 times the size of the SNe~Ia envelope, whereas on the same time a $10^{6}$ G B-field would incur a Larmor radius only 0.001 times the size of the envelope. When the Larmor radius is larger than the envelope, positrons may readily escape. 

\begin{figure}[hbt]
    \centering
    \includegraphics[width=\linewidth]{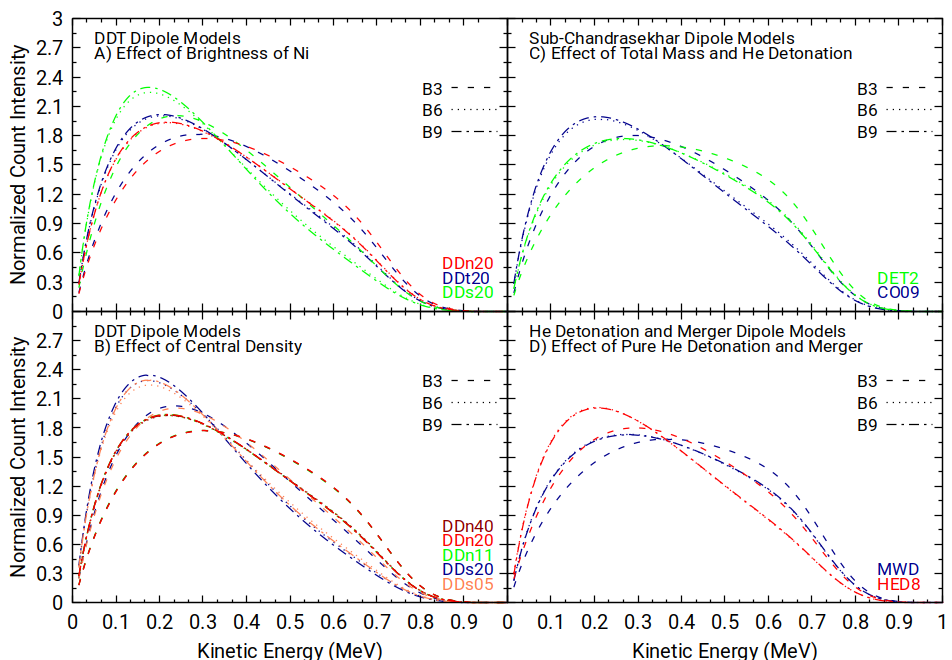}
    \caption{Total integrated escaped positron energy spectrum from day 0 to 2000 for each explosion dipole model. Panel A shows difference between total $^{56}$Ni production (upper left), panel B shows difference between central density (lower left), panel C shows difference between helium detonation masses (upper right), and panel D shows the difference between a merger and a helium detonation (lower right). These models are normalized to unity. The nomenclature used is described in $\S$ \ref{sect:esc_frac}. }
    \label{energy_spectrum_full_dipol}
    \centering
    \includegraphics[width=\linewidth]{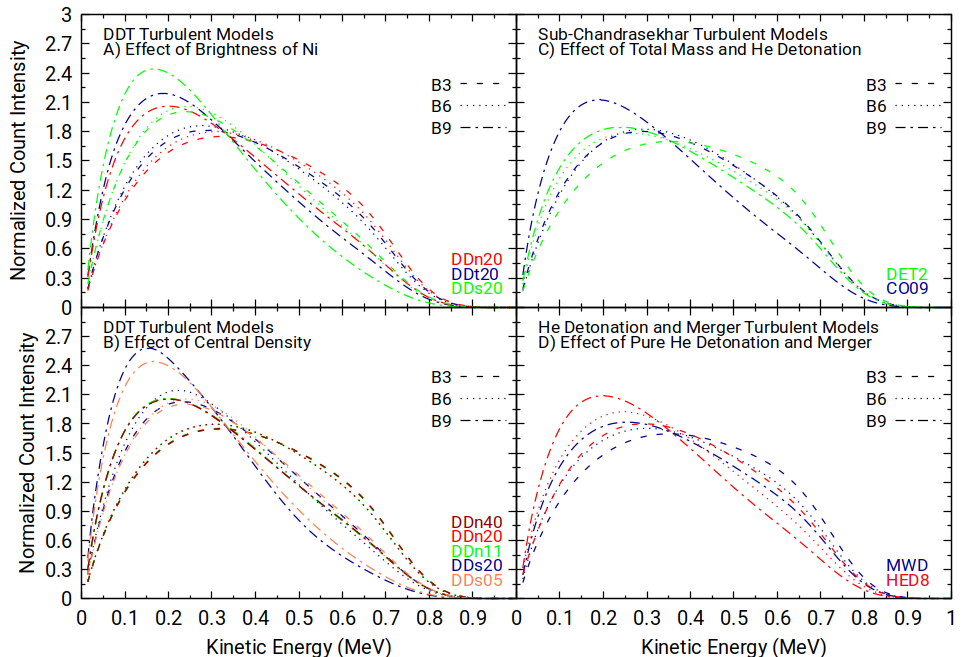}
    \caption{Same as Figure \ref{energy_spectrum_full_dipol}, but with turbulent models.}
    \label{energy_spectrum_full_turb}
\end{figure}


Figure \ref{flux_day_dipol} and \ref{flux_day_turb} show the escaping flux of positrons per day from day 0 to day 1000, for dipole and turbulent models, respectively. The general trend for both dipole and turbulent magnetic fields is that as the magnetic field increases, the peak of escape flux shifts down and towards later times. This is due to the increase of path length and higher chance of annihilating in the envelope. It is also evident that subluminous DDT and low mass He triggered detonations produce a lower flux of escaping positrons due to a smaller yield of $^{56}$Ni for subluminous models and more centrally located $^{56}$Ni for He triggered detonations \newtextc{(see Figure \ref{ni_dist}).}
MWD and DET2 have a high total escaping flux \newtextc{due to their lower mass and mixing, respectively. Note that subluminous sub-$M_{\mathrm{Ch}}$ have comparably low escape because $^{56}$Ni is more centrally concentrated (Figure \ref{ni_dist}).}

Table \ref{tab:tab2} shows the total \newtextc{integrated} positron escape fraction for our various model configurations. We find a range of \newtextc{integrated} positron escape fractions from \newtextc{$<$ 0.01 to 5.99\%.} Subluminous SN~Ia have a lower escape fraction, which is due to a lower yield of $^{56}$Ni primarily produced at lower velocities, meaning the positrons must interact with the envelope more and thus have a higher chance to annihilate locally. For each model, dipole and turbulent, the escape percentage is found relatively the same for low-field models B0 and B3. 
At B-fields $\geq 10^6$ G the escape percentage drops by an approximate factor of \newtextc{three} for dipole fields and by a factor of \newtextc{ten} for turbulent models. \newtextc{B is important and future observations of light curves and spectral profiles will further reduce the uncertainties in its strength and configuration which will help determine the contribution of SNe~Ia to the Galactic positrons. It must also be made clear that $M_{\mathrm{Ch}}$ and sub-$M_{\mathrm{Ch}}$ models have differing total escape fraction ranges, which will lead to alternate contribution rates (see $\S$ \ref{sect:case} and \ref{sect:con}).}

\subsection{Energy Spectra}
\label{sect:ener}

\begin{table*}[t]
\begin{center}
    
\caption{Mean positron energy for each model in MeV. Column one indicates the model, where the preceding columns then give the average positron energy for various dipole or turbulent magnetic field strengths. The nomenclature used can be found in $\S$ \ref{sect:esc_frac}.}
\begin{tabular}{c|cccccccc}
\hline
\hline
  & \multicolumn{2}{c}{B0} & \multicolumn{2}{c}{B3} & \multicolumn{2}{c}{B6} & \multicolumn{2}{c}{B9}  \\ 
 Model& dipole & turb & dipole & turb & dipole & turb & dipole & turb\\ 
 &(MeV)&(MeV)&(MeV)&(MeV)&(MeV)&(MeV)&(MeV)&(MeV)\\\hline
DDs05 &  0.33 & 0.33 & 0.33 & 0.33 & 0.29 & 0.32 & 0.29 &  0.26 \\
DDs20  & 0.33 & 0.33 & 0.33 & 0.33 & 0.29 & 0.28 & 0.29 & 0.28 \\
DDt20  & 0.36 & 0.36 & 0.36 & 0.36 & 0.32 & 0.36 & 0.32 & 0.30  \\
DDn11 & 0.37 & 0.37 & 0.37 & 0.37 & 0.33 & 0.37 & 0.33 & 0.32\\
DDn20  & 0.37 & 0.38 & 0.37 & 0.38 & 0.33 & 0.37 & 0.33 & 0.32  \\
DDn40  & 0.37 & 0.38 & 0.37 & 0.38 & 0.33 & 0.37 & 0.33 & 0.32  \\
DET2  & 0.39 & 0.39 & 0.39 & 0.39 & 0.36 & 0.36 & 0.36 & 0.35  \\
HED8  & 0.36 & 0.36 & 0.36 & 0.36 & 0.32 & 0.34 & 0.32 & 0.31  \\
MWD  & 0.39 & 0.39 & 0.39 & 0.39 & 0.37 & 0.37 & 0.37 & 0.35  \\
CO09  & 0.36 & 0.36 & 0.36 & 0.36 & 0.33 & 0.36 & 0.33 & 0.31  \\ \hline
\end{tabular}
\label{tab:mean-ener}
\end{center}
\end{table*}

The energy spectrum of the escaping positrons is an important outcome of our calculations: this will determine for how long a positron will survive in the interstellar environment of the source, and thus how far it may propagate until eventually annihilating. 
Therefore, in this sub-section we present positron energy spectra from our SN~Ia models, combining the escaping positrons from radioactive decay with high energy positrons from pair-production within the supernova envelope.

Figures \ref{energy_spectrum_full_dipol} and \ref{energy_spectrum_full_turb} show the normalized intensities versus energy for the escaping positrons, for our dipole and turbulent magnetic-field models, respectively. Higher magnetic fields cause the peak of the energy spectrum to shift towards lower energies, and the peak width (measured as full width at half maximum intensity, FWHM) to decrease. This is due to a longer path length of positrons within the source, hence more interactions, and a higher chance to lose energy before their escape. 

The spectra for subluminous models peak at lower energies. This is due to $^{56}$Ni being produced at lower velocities, causing positrons to travel through more material and thus have more chances of losing energy. Despite large early flux for the DET2 and MWD models, their energy spectra remain relatively smooth, an effect caused by the strong centrally-peaked Ni production in these models. 
The turbulent magnetic fields do not shift the peak more, but only slightly increase its strength at lower energies. It is also worth mentioning that the minor secondary peaks around $\approx$0.65 MeV are explained by the high-energy positrons from pair production.

Table \ref{tab:mean-ener} shows the mean energies for each model, where the mean energy amongst all models is $\approx0.35$ MeV. 
The higher magnetic fields, starting at B6, causes the mean energy to drop by approximately 10\%. This reduction is further amplified by an additional 10\% for turbulent fields $\geq$ B9. Subluminous models have a lower mean energy than a normal bright by an approximate 10\% difference, due to the reasoning stated above.  

\subsection{Case Studies on Galactic SNe~Ia Class Population}
\label{sect:case}
As discussed in $\S$ \ref{sec:Introduction}, the progenitor system of SNe~Ia can be differentiated into three classes: subluminous, transitional, and normal bright, each of which are described by their light curve. A large majority of SNe~Ia are classified into one of these categories without having enough observational constraints to identify its explosion scenario. Therefore, in this section we want to explore the observational information from each SN~Ia class, and compare them to case studies that involve different model variations of DDTs and He triggered detonations, in order to determine if the explosion mechanism in SNe~Ia carries some information forward to the population and observational appearance of Galactic positrons.
We also want to determine a range of positron yields from SNe~Ia that can be used in future Galactic positron studies.

\begin{table*}[t]
\centering
0\caption{Total averaged escaped positrons for case one, two, three, and four for B0, B6, and B9 weighted by the ratio from CSPI, CSPII. \newtextc{In parentheses, we give the escape fraction in \%}. A description of the models used in each case study can be found in $\S$ \ref{sect:case}.}
\newtextc{\begin{tabular}{c|ccccccc}

\hline
\hline
 &  & \multicolumn{2}{c}{CSPI}         & \multicolumn{2}{c}{CSPII}    & \multicolumn{2}{c}{CSPI/II}   \\ 
&  B-field & dipole&  turb & dipole & turb &   dipole & turb \\
Case  & ($10^{\#}$G)      & ($10^{52}$ e$^{+}$)(\%) & ($10^{52}$ e$^{+}$)(\%) & ($10^{52}$ e$^{+}$)(\%) & ($10^{52}$ e$^{+}$)(\%)   & ($10^{52}$ e$^{+}$)(\%) & ($10^{52}$ e$^{+}$)(\%)    \\\cline{2-8}
One  & B0       & 5.40 (2.80) & 5.26 (2.73) & 6.58 (2.94)& 6.42 (2.87) & 6.03 (2.88)& 5.88 (2.80)\\
&  B6       & 2.00 (1.04) & 0.31 (0.16)& 2.45 (1.09)& 0.38 (0.17)& 2.24 (1.07)& 0.35 (0.17)\\
&  B9       & 1.96 (1.02)& 0.16 (0.08)& 2.41 (1.07)& 0.19 (0.08)& 2.19 (1.05)& 0.18 (0.08)\\\hline \hline
 &  & \multicolumn{2}{c}{CSPI}         & \multicolumn{2}{c}{CSPII}     & \multicolumn{2}{c}{CSPI/II}     \\ 
&  B-field & dipole  & turb  & dipole & turb & dipole & turb  \\
Case  & ($10^{\#}$G)      & ($10^{52}$ e$^{+}$)(\%)& ($10^{52}$ e$^{+}$)(\%)& ($10^{52}$ e$^{+}$)(\%)& ($10^{52}$ e$^{+}$)(\%)  & ($10^{52}$ e$^{+}$)(\%)& ($10^{52}$ e$^{+}$)(\%)       \\\cline{2-8}
Two  & B0       & 10.9 (5.28)& 10.9 (5.26)& 13.1 (5.63)& 13.0 (5.61)& 12.1 (5.48)& 12.1 (5.47)\\
&  B6       & 8.55 (4.11)& 1.38 (0.66)& 10.5 (4.49)& 1.71 (0.73)& 9.58 (4.34)& 1.56 (0.70)\\
&  B9       & 5.94 (2.86)& 0.66 (0.32)& 7.23 (3.10)& 0.82 (0.35)& 6.63 (3.00)& 0.75 (0.34)\\\hline \hline
 &  & \multicolumn{2}{c}{CSPI}         & \multicolumn{2}{c}{CSPII}     & \multicolumn{2}{c}{CSPI/II}     \\ 
&  B-field & dipole  & turb  & dipole & turb & dipole & turb  \\
Case  & ($10^{\#}$G)      & ($10^{52}$ e$^{+}$)(\%)& ($10^{52}$ e$^{+}$)(\%)& ($10^{52}$ e$^{+}$)(\%)& ($10^{52}$ e$^{+}$)(\%)  & ($10^{52}$ e$^{+}$)(\%)& ($10^{52}$ e$^{+}$)(\%)      \\\cline{2-8}
Three  & B0       & 5.84 (2.87)& 5.71 (2.80)& 6.91 (2.97)& 6.73 (2.90)& 6.39 (2.93)& 6.24 (2.86)\\
&  B6       & 2.16 (1.06)& 0.31 (0.15)& 2.57 (1.10)& 0.39 (0.17)& 2.37 (1.08)& 0.35 (0.16)\\
&  B9       & 2.11 (1.03)& 0.15 (0.07)& 2.51 (1.08)& 0.19 (0.08)& 2.33 (1.06)& 0.18 (0.08)\\\hline \hline
 &  & \multicolumn{2}{c}{CSPI}         & \multicolumn{2}{c}{CSPII}    & \multicolumn{2}{c}{CSPI/II}      \\ 
&  B-field & dipole  & turb  & dipole & turb  & dipole & turb \\
Case  & ($10^{\#}$G)      & ($10^{52}$ e$^{+}$)(\%)& ($10^{52}$ e$^{+}$)(\%)& ($10^{52}$ e$^{+}$)(\%)& ($10^{52}$ e$^{+}$)(\%)  & ($10^{52}$ e$^{+}$)(\%)& ($10^{52}$ e$^{+}$)(\%)       \\\cline{2-8}
Four  & B0       & 10.5 (5.35)& 10.5 (5.33)& 12.8 (5.68)& 12.8 (5.66)& 11.7 (5.54)& 11.7 (5.53)\\
&  B6       & 8.39 (4.26)& 1.38 (0.70)& 10.3 (4.60)& 1.71 (0.75)& 9.45 (4.46)& 1.56 (0.73)\\
&  B9       & 5.79 (2.93)& 0.66 (0.33)& 7.12 (3.16)& 0.82 (0.36)& 6.50 (3.07)& 0.75 (0.35)\\\hline 
\end{tabular}}
\label{tab:case1}

\end{table*}

A key property in interpreting the light curve of SNe~Ia is the Phillips relation, i.e., the difference between the brightness magnitude at the time of maximum brightness and the brightness 15 days later. This difference, labeled $\Delta m_{15,s}$, is what has been shown to differentiate between various progenitor/explosion systems \citep{Taubenberger:2017}.  For normally-bright SNe~Ia, $\Delta m_{15,s}(V)$ \newtextc{$<$} 0.85, transitional from 0.85 to 1.2, and for subluminous events beyond 1.2, corresponding to $\Delta m_{15,s}(B)$ of 1.3, 1.7 and 1.7+, respectively.
\newtextc{Equivalent to $\Delta m_{15,s}$,  $s_{B-V}$ is commonly used for characterizing SNeIa light curves. It is defined as time of the maximum (B-V) color relative to B maximum divided by 30 days. The transition between normal-bright and subluminous SNe~Ia corresponds to $s_{B-V}$ of 0.7 and 0.5, respectively \citep{burns_2014}.}

The Carnegie Supernova Project (CSP) has produced two data sets that give the properties of SNe~Ia within the Galaxy; from the first dataset, CSP-I, we obtained a normal, transitional, and subluminous event rate ratio of 1:0.2:0.28,
respectively. From CSP-II we obtain a ratio of  1:0.11:0.075. The difference in values is due to selection effects in the surveys, which take into account either local or high-z SNe~Ia, respectively; the combined ratios are 1:0.14:0.15.
Normally-bright supernovae will play the dominant role in the positron injection from SNe~Ia.\footnote{\newtextc{$^{56}$Ni is produced at high densities in nuclear statistical equilibrium. 
Less $^{56}$Ni in sub-luminous SNe~Ia means the majority of $^{56}$Ni is at larger optical depths resulting in lower positron escape (Figure \ref{ni_dist}).
In principle, the scheme can be broken if the majority of $^{56}$Ni is produced in outer He-layers that have not yet been observed (see \citet{Hoeflich2017book}, and references therein).}}
With this understanding, we can select appropriate models, and create rate-weighted integrated energy spectra and total fluxes of SNe~Ia positrons. We will put our focus to magnetic field strengths of B0, B6, and B9, as the properties of model B3 are very similar to model B0, and those of models B11 and B13 to model B9 \newtextc{(see $\S$ \ref{sect:esc_frac})}. We identify four different cases that we consider indicative of the different types of progenitors systems. Each case is described below, and the integrated number of escaping positrons can be found in Table \ref{tab:case1}.\par
\begin{figure}[b]
    \centering
    \includegraphics[width=\linewidth]{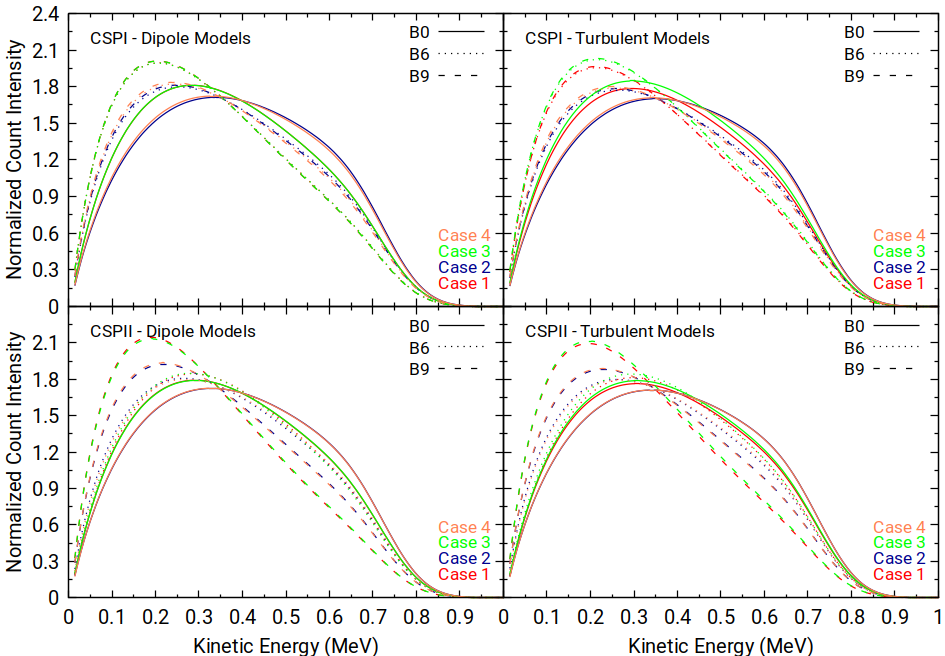}
    \caption{Average energy spectra for case 1, 2, 3, and 4, for dipole (left) and turbulent (right) magnetic-field models B0, B6, and B9. The upper two panels are representative of the CSP-I and the lower two are representative of the CSP-II event type ratios. Each curve is normalized to unity. A description of the models used in each case study can be found in $\S$ \ref{sect:case}. }
    \label{cas_ener_spe}
\end{figure}

\textbf{Case 1}: This case concerns delayed detonation models, which have been presented in panel A for most of the figures above. This selection has DDn20 as the normal bright, DDt20 as the transitional, and DDs20 as the subluminous  SN~Ia model. This case is shown as the red curves in Figure \ref{cas_ener_spe}.

\textbf{Case 2}: These are the helium detonation models. For this selection, DET2 is the normal bright, HED8 is the transitional, and CO09 is the subluminous event type. This case is shown as the blue curves in Figure \ref{cas_ener_spe}.

\textbf{Case 3}: A combined case of the delayed and helium detonation models, with an emphasis on the delayed detonation variant. DDn20 will be the normal bright, HED8 will be the transitional, and DDs20 will be the subluminous model type. This case is shown as the green curves in Figure \ref{cas_ener_spe}. 

\begin{table*}[t]
\begin{center}

\caption{Mean energy, in MeV, for Case 1, 2, 3 and 4 for B fields of B0, B6, and B9, averaged between CSPI and CSPII. A description of the models used in each case study can be found in $\S$ \ref{sect:case}.}
\begin{tabular}{c|cccccccc}
\hline
\hline
 & \multicolumn{2}{c}{Case 1}  & \multicolumn{2}{c}{Case 2} & \multicolumn{2}{c}{Case 3}   & \multicolumn{2}{c}{Case 4}  \\ 
    B Field       & dipole & turb & dipole & turb & dipole & turb & dipole & turb   \\
   ($10^{\#}$G) & (MeV) & (MeV) & (MeV) & (MeV) & (MeV) & (MeV) & (MeV) & (MeV)   \\\hline
B0       & 0.36 & 0.36& 0.38&0.38 &0.36 &0.37 & 0.38&0.38   \\
B6       & 0.32& 0.35& 0.35& 0.35& 0.32& 0.35& 0.35&0.35   \\
B9       & 0.32& 0.30& 0.35&0.34 & 0.32& 0.31& 0.35& 0.34 \\\hline
\end{tabular}
\label{tab:cas_mean}
\end{center}
\end{table*}

\textbf{Case 4}: Same as case 3, but with an emphasis on helium detonations. DET2 will be the normal bright, DDt20 will be the transitional, and CO09 will be the subluminous model. This case is shown as the coral curves in Figure \ref{cas_ener_spe}. 

Figure \ref{cas_ener_spe} shows the energy spectra for the four case studies, weighted by the ratios from CSP-I and CSP-II, for dipole and turbulent magnetic fields of our B0, B6 and B9 model variants. As the magnetic field increases, the peak shifts to lower energies, due to the increase in positron path length and a higher chance of losing energy (see $\S$ \ref{sect:ener}). 
Table \ref{tab:cas_mean} shows the mean energy for the four case models averaged between CSPI and CSPII. The difference in mean energy between the four cases is less than 10\%. The difference in mean energy between B0 and B9 is approximately 10\%, with an additional 10\% for turbulent models.

The variations between the four cases show remarkable patterns and stability, which leads us to suggest that the explosion mechanism may be unimportant and not discriminative for the energy spectra of the escaping positrons, \newtextc{but that the energy shift is dominated by the B-field strength.}

\newtextc{From Table \ref{tab:case1}, we find that as B changes from low to high strengths the positron escape 
fraction decreases by factors of $\approx 2$ and $\approx 24$ for dipole and turbulent morphologies, respectively. The difference between dissimilar cases carries an additional factor of $2-4$ difference, depending on the size and morphology of B, where the reason for the variation is mostly due to the difference in explosion scenarios for normal-bright SNe~Ia (see Figures \ref{flux_day_dipol} and \ref{flux_day_turb}). 

 Following \citet{hristov_2021} and as discussed in $\S$ \ref{sec:Introduction} and below in $\S$ \ref{sect:con}, we expect $M_{\mathrm{Ch}}$ explosions to have a turbulent field morphology with a lower strength estimate of $\approx 10^{6}$ G. This would emphasize the escape fractions of case one and three to take on a range from 0.07 to 0.17\%.  Sub-$M_{\mathrm{Ch}}$ explosions may have either a turbulent or dipole field, where the lower dipole B-field estimate is $\approx 10^{9}$ G. Thus, the escape fractions for cases two and four would range from 0.32 to 3.16 with or 0.75\% without dipole field morphologies, respectively.
The differences in the positron injection into the Galaxy will mostly depend on
our knowledge of whether $M_{\mathrm{Ch}}$ or sub-$M_{\mathrm{Ch}}$ dominate normal bright 
events and on the strength and morphology of B, but we highlight that an even greater source of uncertainty is in the SNe Ia rate, where upcoming SNe Ia surveys will largely constrain all of these channels (see $\S$ \ref{sect:con}).}



\section{Discussion}
\label{sect:con}

In $\S$ \ref{test} and $\S$ \ref{sect:ener}, we presented positron escape fractions and energy spectra for a wide variety of SNe~Ia explosion scenarios that include subluminous, transitional, and normal bright events for DDTs, He triggered detonations, and merging explosion mechanisms. We tested either dipole or small-scale turbulent structures, that range in strength from 1 to $10^{13}$ G. Then, in $\S$ \ref{sect:case}, we explored four cases of  explosion models, weighing outcomes by their observed SN~Ia class ratio from CSP-I/II. 

A natural question that arises from our study is: what are the physical origins of our magnetic-field discriminating results?

Recent analyses have estimated lower \newtextc{magnetic field limits for SNe~Ia of $10^{6}$ and $10^{9}$ G, for turbulent and dipole B-fields, respectively} \citep{tiara15, hristov_2021}. Even though several WDs have been directly observed to have magnetic fields of $\approx 10^{7}$ G, the majority of WD fields are small \citep{liebert03,schmidt03,Silvestri07,tout08}. This discrepancy appears to require field amplification during some phase \newtextc{during the progenitor evolution or the explosion}. 

Dynamo theory \citep{Brandenburg05} provides mechanisms that can amplify field strengths up to a saturation strength of $\approx 10^{14}$ G \citep{hristov_2018}, where the scale and morphology of the B-field depend on length scale of plasma flow and phase during amplification.

\begin{table*}[t]
\begin{center}

\caption{Estimated positron annihilation rate from SNe~Ia in $e^{+}s^{-1}$ for case 1, 2, 3, 4 and B fields of B0, B6, and B9, averaged between CSPI and CSPII. Wen use a value of 0.54 SNe~Ia per century. A description of the models used in each case study can be found in $\S$ \ref{sect:case}.}
\newtextc{\begin{tabular}{c|cccccccc}
\hline
\hline
 & \multicolumn{2}{c}{Case 1}  & \multicolumn{2}{c}{Case 2} & \multicolumn{2}{c}{Case 3}   & \multicolumn{2}{c}{Case 4}  \\ 
    B Field       & dipole & turb & dipole & turb & dipole & turb & dipole & turb   \\
   ($10^{\#}$G) & ($10^{42}~e^{+}/s$) & ($10^{42}~e^{+}/s$) & ($10^{42}~e^{+}/s$) & ($10^{42}~e^{+}/s$) & ($10^{42}~e^{+}/s$) & ($10^{42}~e^{+}/s$) & ($10^{42}~e^{+}/s$) & ($10^{42}~e^{+}/s$)   \\\hline
B0       & 10.3 & 10.1 & 20.7 & 20.7 & 10.9 & 10.7 & 20.1 & 20.0   \\
B6       & 3.84 & 0.60 & 16.4 & 2.67 & 4.07 & 0.60 & 16.2 & 2.66  \\
B9       & 3.76 & 0.30 & 11.3 & 1.28 & 3.98 & 0.30 & 11.1 & 1.28\\\hline
\end{tabular}}
\label{tab:injection}

\end{center}
\end{table*}

\newtextc{ In all scenarios, this amplification may occur 
during the accretion phase. The accretion phase will induce meridional circulation in the rotating WD \citep{pranab_2017}}, which may amplify initial B-fields within scales equivalent to the radius of the WD, leading to large scale dipole structures.   For dynamical mergers, high turbulent magnetic fields may develop, depending on the dynamical merging/colliding process; e.g., amplification may be caused by differential rotation within a common envelope \citep{tout08}, or amplification may be caused by shear stress during the initial merging.\newtextc{ For He-triggered detonation, the development of turbulent morphologies cannot be ruled out. Thus, 
we may find dipole or turbulent B-field morphologies in
sub-$M_{Ch}$ and merger scenarios.
 In $M_{Ch}$ explosions, large B-fields may also be produced} during the late stage up to the deflagration, the so called 'smoldering phase' \citep{hoeflich_stein_2002, zingale11}. The central convective motion during the smoldering phase may amplify fields on scales comparable to the pressure scale height in the initial WD, i.e. $\approx 2$ to $5 \% $ the size of the WD \citep{hoeflich_stein_2002}. \newtextc{In principle, high B fields may be produced during the deflagration phase by RT-instabilities \citep{khokhlov95, niemeyer95}, though, this mechanism has been found to be ineffective \citep{hristov_2021}. However, independent from the field morphology at the time of the runaway, }
 Rayleigh Taylor instabilities may twist these newly amplified B-fields on scales equivalent to the Rayleigh Taylor scale height, i.e. $\approx 5$ to $10 \% $ the size of the WD \citep{hristov_2021}. This will form into a more turbulent magnetic-field structure \newtextc{independent from the origin of the high-fields.}

The magnetic field will then be imprinted onto the SNe~Ia envelope, and the homologous expansion will imply very limited morphological change: the magnetic field will be preserved in its dipole or turbulent configuration, with the strength being dependent on the number of dynamo e-foldings. Positrons will then follow these magnetic field lines, where their local positioning is dictated by the ratio between their Larmor radius and the eddie sizes within the magnetic field. This argument differs from the one made by \citet{chan_ligenfelter_1993} and \citet{milne_1999_true, milne_2001}, who assumed that positrons will follow weak and radially combed-out magnetic field lines, or that positrons will be completely trapped by turbulent fields and carried out with the expanding local material.

Within this understanding, we were then able to develop SNe~Ia models with positron transport solutions that tested different magnetic field configurations (see $\S$ \ref{sect:esc_frac}). \newtextc{Over the entire range of models}, we found positron escape fractions from \newtextc{ $<$0.01 to 5.99\%}, a shift in mean energy from $\approx 0.6$ to $\approx 0.35$ MeV, and typical initial positron escape 150 days post-explosion. \newtextc{Though somewhat lower because of the field configurations, our results are consistent with} the results of \citet{chan_ligenfelter_1993}, who found a range of escape fractions from 0.08 to 15\% and a shift in mean energy to $\approx0.4$ MeV. \newtextc{However, for the conclusions,} it differs from the light curve studies of \citet{milne_1999_true, milne_2001}, who found indications of initial positron escape 95 to 180 days after the explosion, escape fractions from 0.0 to 11.3\%, \newtextc{and proposed radial B-fields based on the flawed assumption that the optical light curves can be used as proxy for the bolometric light curve, as discussed in $\S$ \ref{sec:Introduction}}.

   Although we follow a similar integration method as \citet{chan_ligenfelter_1993} and \citet{milne_1999_true, milne_2001} for \newtextc{low strength} dipole fields, our path lengths are increased because of directional dependence until their escape at the poles, instead of being radially beamed out in all directions. Additionally, we implemented fully explicit integration methods for \newtextc{intermediate to high strength} dipole fields and all turbulent field strengths, which is required because of the tighter gyration, but also due to magnetic field scale structures allowing positrons to escape its local environment as their Larmor radius becomes larger than the size of the eddie. 
We must also emphasize that the inclusion of transitional and subluminous models will produce \newtextc{more centrally located }$^{56}$Ni (see Figure \ref{ni_dist}), meaning positrons will have an additional region of envelope to traverse through, which will naturally lead to lower escape fractions and mean energies.

\newtextc{ As discussed in $\S$ \ref{sect:case}, we significantly narrowed the uncertainties in the positron escape for given explosion scenarios based on the observed Galactic SNe~Ia population ratio. 
If we translate these findings to the congruent positron yields, then we find positron emission per average SNe~Ia for $M_{\mathrm{Ch}}$ dominated explosions to extend up to $0.35\times 10^{52}$ $e^{+}$. For populations with sub-$M_{Ch}$ explosions dominating, we find emissions of 1.56 and 6.63$\times 10^{52}$ $e^{+}$, for turbulent and dipole field morphologies, respectively.} 

\newtextc{If we take the commonly used SNe~Ia rate of 0.54 per century \citep{li_2011}, then $M_{\mathrm{Ch}}$ explosions will contribute with a rate upwards of $0.6 \times 10^{42}$ $e^{+}$s$^{-1}$, whereas sub-$M_{\mathrm{Ch}}$ explosions would contribute with rates that reach 2.67 or 11.3$\times 10^{42}$ $e^{+}$s$^{-1}$, for turbulent or dipole fields respectively (see Table \ref{tab:injection}).}
From the Galactic positron annihilation value of ($4.9 \pm 1.7) \times 10^{43}$ $e^{+}$s$^{-1}$ found in \citet{Siegert2017PhD}, this would make \newtextc{$M_{\mathrm{Ch}}$} accountable for \newtextc{0.012} of the total Galactic positron annihilation rate, \newtextc{whereas sub-$M_{\mathrm{Ch}}$ would account upwards of 0.054 or 0.23 of the total Galactic positron annihilation rate, for turbulent or dipole fields, respectively. These estimates are largely} under the assumption of steady state between injection and annihilation. \newtextc{To first order, the SNe~Ia contribution is smaller by a factor of 5 to 100 times than needed. Note, however, that better estimates require detailed simulations 
of the Galactic positron annihilation with the revised SNe~Ia contributions.
We emphasize that our results sensitively scale with the SNe~Ia rate (e.g. if we were to take the SNe~Ia rate of 0.21 per century from \citet{cappellaro_1999}, then our estimates would drop by an additional factor of $\approx  2$). Regardless, we find that SNe~Ia are not the dominant contributors to the Galactic positron whereas most previous estimates range from  $40 $ to more than 99 \%}
(see e.g. \citet{chan_ligenfelter_1993, purcell_1997, milne_1999_true, milne_2001, higdon_2009, prantzos_2011, martin_2012,alexis_2014,  Siegert2017PhD}).
Differences in these values result from: variations in assessments of the observed Galactic positron annihilation rate, lower escape fractions, the use of different progenitor magnetic field structures, considerations from the observed ratio of subluminous, transitional and normal bright SNe~Ia in the context of various explosion scenarios, \newtextc{and the SNe~Ia rate.} 
Although this may be a striking conclusion, this reduction falls within the uncertainty limits of other sources, so that X-ray binaries, pulsars, and cosmic rays could recover this lost fraction of positrons annihilating in the Galaxy \citep{Siegert2017PhD}. These alternate source contributions are, however, also mostly theoretical estimates based on sparse observational constraints and additional assumptions. More solid conclusions would require additional observations and analysis.


\section{Conclusion}
\label{sect:conclusion}
We have explored in great detail the positron channel for a wide range of SNe~Ia scenarios and the influence they may have on the population of annihilating positrons in our Galaxy. \newtextc{Using population synthesis, we identified normal-bright SNe~Ia as the main positron contributors from SNe~Ia. The positron energy distribution hardly depends on the explosion scenario. Though variations within the class of explosions is small (see $\S$ \ref{sect:case}), the positron injection rate depends on the dominant explosion mechanism, the initial size and morphology of the B-field in the WD (see Table \ref{tab:case1}), and the SNe~rate (see $\S$ \ref{sect:con} and Table \ref{tab:injection}). 
Regardless, it's unlikely that SNe~Ia are a major contributor to the population of Galactic positrons.}

\newtextc{Most of the uncertainties will be largely reduced by ongoing and upcoming observations.
For example, the SNe-rate will be greatly improved by the Large Synoptic Survey Telescope \citep{LSST} and WFIRST \citep{WFIRST}. Additionally, distinguishing explosion scenarios based on their signatures from the outermost and central layers can be accomplished with transient surveys for early detections such as: ZTF \citep{sharon22, yao19}, ATLAS \citep{ATLAS}, ASASSN \citep{shappee14, kochanek17}, SPECPOL \citep{yi20}, POISE \citep{POISE}, in combination with late-time nebular spectra in the NIR and MIR with VLT, Keck, and JWST (e.g., \citet{graham17,Galbany2019,jacobson2020,ashall21A,hoeflich2021}). Inevitably, the number of bolometric light curves between 200-800 days and late-time nebular will become increasingly significant, allowing for tighter constrains on progenitor B-fields (see \S \ref{sec:Introduction}).}



We caution, however, that there are significant limitations of our study. 
The models discussed here assume spherical symmetry. Our models predict insignificant chemical mixing. If such mixing is considered, this may slightly increase the escape of positrons and push the energy spectrum peak to higher energies, as more $^{56}$Ni would be found at higher velocities. Magnetic fields may curb this effect \newtextc{\citep{hristov_2018}, 
but more detailed multidimensional simulations will be studied in the future.}

In our work, we have constrained the importance of positrons from SNe~Ia for our Galaxy. But currently we are limited in the sophistication and realism of our Galactic modeling capabilities. To get a better description of the spatial distribution of positrons from SNe~Ia, new detailed simulations on the spatial distribution of SNe~Ia, and on  Galactic positron transport that includes our positron yields and energy spectra, along with a more precise SNe~Ia rate, will hopefully help lead to a resolution of the positron puzzle.

\section*{Acknowledgements}

\newtextc{ We thank the referee for carefully reading the manuscript and, in particular, pointing out an inconsistency in time-integrated positron escape fraction of Table 1 unveiling a problem with our time-integration routine.}
We acknowledge support by National Science
Foundation (NSF) grant AST-1715133.

\facility{The simulations have been performed on the computer cluster
  of the astro-group at Florida State University.}

\software{ \hydra\ \citep{hoeflich2003hydra,hoeflich_2009_hyro,hoeflich2017,hristov_2021}, and the plotting software Gnuplot version 5.4 by T. Williams and C. Kelley \& contributors (http://sourceforge.net/projects/gnuplot).}


\begin{sidewaystable}[htb!]
\begin{center}
\caption{Properties of our models. The nomenclature used for each model is described in Sect. \ref{sect:esc_frac}.
We give the main sequence mass $M_{\rm MS}$ (in $M_{\odot}$), metallicity $Z$ (in $Z_{\odot}$), central density $\rho_c$ (in $10^9$gcm$^{-3}$), transition density $\rho_{\rm tr}$ (in 
$10^6$gcm$^{-3}$), and the resulting total $^{56}Ni$ production $M_{\rm Ni}$ (in $M_{\odot}$). 
For the HeD-models, we assume a $C/O=1$ core plus a He-shell with solar metallicities. 
The optical properties, decline rate $\Delta m_{15,s}$, peak magnitudes $M_{B,V}$, color and rise time at $V$-band maximum $(B-V)_{\rm max}$ and $t_s$ are given in the natural system of the Swope and du Pont telescopes. In addition, we give the color-stretch $sbv$ taking into account the model uncertainties in $t_{max}$ and the maximum bolometric luminosity $L_{bol}$ which includes the UV. For the uncertainties of the observables, see the corresponding references given in the last column. Note that the double-detonation are based on low-resolution calculations with 274 depth points. (+) Three models with $\rho_c $ of $1.1$, $2$, and $4\times10^9$~gcm$^{-3}$; and $M_{\rm Ni}$ of $0.61$, $0.56$, and $0.55$$ ~M_{\odot}$, respectively.}
\begin{tabular}{|c|c|c|c|c|c|c|c||c|c|c|c|c|c|c||c|}
	\hline Model(alias) & $M_{\rm MS}$& $Z$ & $M_{\rm WD}$ &  $\rho_c$ &$ \rho_{\rm tr}$ & $M_{\rm Ni}$  & $M_V$ &$\Delta m_{15,s}(V)$ &  $M_B$ & $\Delta m_{15,s}(B)$ &  $(B-V)_{\rm max}$ &  $t_s$ & $sbv$  & $log(L_{bol})$ &  Refs. \\
\hline 
\hline DDs05 & 5. & 1. &  $M_{Ch}$ & 0.5 &  8.  &  0.188   &   -17.63  & 1.39 & -17.24   & 1.93 & 0.41 & 15.26  & 0.29-0.53  & 42.70 &  (1)      \\
\hline DDs20 & 5. & 1. &$M_{Ch}$ & 2. &  8. &  0.095  & -17.22  & 1.41 & -16.76  & 1.99 & 0.487 & 14.38  & 0.33-0.60 & 42.21 &  (1)     \\
\hline DDt20 & 5. & 1. & $M_{Ch}$ & 2. &  16. &  0.268    &  -18.37  & 1.18 & -18.22  & 1.67 & 0.157 & 15.37  & 0.50-0.67   & 42.84 & (1)   \\
\hline DDn($+$)  &  5. & 1. & $M_{Ch}$ & ($+$) & 27. &  ($+$) &     -19.35  & 0.64 & -19.38  & 0.95 & -.003 & 18.80  & 0.99-1.08  & 43.25 & (1)   \\
\hline HeD8 & n/a & 1. &0.8+0.16  &   0.025& n/a&  0.526  &   -19.21  &  1.38 & -19.23  & 1.33  & -0.02 & 13.8 & n/a & 43.22 &  (2)  \\
\hline CO09 & n/a & 1. &0.9  &   0.016& n/a&  0.18  &   -18.61  &  1.27 & -18.24  & 1.53  & 0.37 & 14.2 & n/a & 42.82 &  (2)  \\
\hline DET2& n/a & 1. &1.2  &   0.004& n/a&  0.63  &   -19.67  &  1.32 & -19.53  & 1.32  & 0.14 & 14.2 & n/a & 43.32 &  (2)  \\
\hline MWD& n/a & 1. &1.6  &   -& n/a&  0.61  &   -19.28  &  0.58 & -19.41  & 0.69  & 0.14 & 19.5 & n/a & 43.27 &  (2)  \\
\hline 
\end{tabular}
\label{table_mod}
\end{center}

{\bf References:} $(1)$ \citet{hoeflich2017} ~
$(2)$ \citet{hk96} 
\end{sidewaystable}

\clearpage

\end{document}